\documentclass[pdflatex,sn-basic]{sn-jnl}

\usepackage{graphicx}%
\usepackage{multirow}%
\usepackage{amsmath,amssymb,amsfonts}%
\usepackage{amsthm}%
\usepackage{mathrsfs}%
\usepackage[title]{appendix}%
\usepackage{xcolor}%
\usepackage{textcomp}%
\usepackage{manyfoot}%
\usepackage{booktabs}%
\usepackage{algorithm}%
\usepackage{algorithmicx}%
\usepackage{algpseudocode}%
\usepackage{listings}%

\usepackage{booktabs}    
\usepackage{graphicx}    
\usepackage{float}       

\usepackage{graphicx}  
\usepackage{float}     

\theoremstyle{thmstyleone}%
\theoremstyle{thmstyletwo}%

\theoremstyle{thmstylethree}%

\begin{document}

\title[Predicting Chest Radiograph Findings from Electrocardiograms Using Interpretable Machine Learning]{Predicting Chest Radiograph Findings from Electrocardiograms Using Interpretable Machine Learning}


\author[1]{\fnm{Julia} \sur{Matejas}}\email{julia.matejas@uol.de}
\equalcont{These authors contributed equally to this work.}

\author[1]{\fnm{Olaf} \sur{Żurawski}}\email{olaf.zurawski@uol.de}
\equalcont{These authors contributed equally to this work.}

\author*[1]{\fnm{Nils} \sur{Strodthoff}}\email{nils.strodthoff@uol.de}

\author*[1]{\fnm{Juan Miguel} \sur{Lopez Alcaraz}}\email{juan.lopez.alcaraz@uol.de}

\affil[1]{\orgdiv{AI4Health Division}, \orgname{Carl von Ossietzky Universität}, \orgaddress{\city{Oldenburg}, \country{Germany}}}


\abstract{
\textbf{Purpose:} Chest X-rays are essential for diagnosing pulmonary conditions, but limited access in resource-constrained settings can delay timely diagnosis. Electrocardiograms (ECGs), in contrast, are widely available, non-invasive, and often acquired earlier in clinical workflows. This study aims to assess whether ECG features and patient demographics can predict chest radiograph findings using an interpretable machine learning approach.

\textbf{Methods:} Using the MIMIC-IV database, Extreme Gradient Boosting (XGBoost) classifiers were trained to predict diverse chest radiograph findings from ECG-derived features and demographic variables. Recursive feature elimination was performed independently for each target to identify the most predictive features. Model performance was evaluated using the area under the receiver operating characteristic curve (AUROC) with bootstrapped 95\% confidence intervals. Shapley Additive Explanations (SHAP) were applied to interpret feature contributions.

\textbf{Results:} Models successfully predicted multiple chest radiograph findings with varying accuracy. Feature selection tailored predictors to each target, and including demographic variables consistently improved performance. SHAP analysis revealed clinically meaningful contributions from ECG features to radiographic predictions.

\textbf{Conclusion:} ECG-derived features combined with patient demographics can serve as a proxy for certain chest radiograph findings, enabling early triage or pre-screening in settings where radiographic imaging is limited. Interpretable machine learning demonstrates potential to support radiology workflows and improve patient care.

}

\keywords{Chest X-ray prediction, Electrocardiogram (ECG), Interpretable machine learning, Multilabel classification, Early triage}


\maketitle

\section{Introduction}\label{sec1}
Pulmonary diseases remain a major global health burden, contributing substantially to morbidity and mortality \citep{ranieri2012ards}, where timely diagnosis and intervention are critical to improving patient outcomes \citep{cosgrove2018barriers}. Chest X-rays (CXRs) are the standard imaging modality for detecting and monitoring pulmonary conditions, providing a non-invasive and widely accessible method to visualize lung structures. However, access to CXR imaging can be limited by logistical constraints, cost, and the need for expert interpretation, particularly in emergency or resource-limited settings \citep{philipsen2015automated}. Technical challenges in CXR interpretation, including label imbalance and inter-observer variability \citep{holste2024towards}, may further delay clinical decision-making, underscoring the need for alternative, rapid, and accessible diagnostic strategies.

Electrocardiograms (ECGs), traditionally employed to diagnose cardiac conditions such as arrhythmias and myocardial infarction \citep{hancock2009aha}, are inexpensive, non-invasive, widely available, and often collected as part of routine assessment during initial clinical evaluation \citep{macallan1990electrocardiogram}. Their portability and ease of acquisition make them particularly attractive in settings where access to advanced imaging modalities such as echocardiography or chest radiography may be constrained \citep{bansal2018portable}. Beyond their primary role in cardiovascular diagnostics, ECGs provide continuous, high-resolution measurements of cardiac electrical activity, offering rich temporal signals that can reflect both acute pathological events and more subtle, long-term physiological alterations \citep{tikkanen2009long}.

Emerging evidence suggests that ECG signals can also reflect systemic physiological changes linked to pulmonary diseases \citep{han2007pulmonary}, positioning them as a promising yet underutilized tool for early triage, diagnosis, and continuous monitoring. Prior studies have examined the relationship between ECGs and pulmonary conditions, either by using ECG features with traditional machine learning (ML) models to diagnose a limited set of diseases \citep{lopez2024estimation}, or by relying on echocardiograms for specific conditions such as pulmonary hypertension \citep{roberts2011diagnosis}. While these approaches show potential, prior work using ECG features for pulmonary disease classification has typically relied on qualitatively different labels such as discharge diagnoses \citep{lopez2024estimation}, which, although clinically meaningful, represent coarser and less direct outcomes compared to image-based findings available immediately after acquisition and less prone to future confounding. Other studies have relied on echocardiograms for a single condition such as pulmonary hypertension, often using less interpretable model architectures \citep{roberts2011diagnosis}. In contrast, our work leverages ECGs to predict a wide spectrum of direct chest X-ray findings, providing more clinically aligned targets and enhancing interpretability to foster clinical trust.


In this study, we propose an interpretable machine learning framework to predict chest X-ray (CXR) findings using only structured features from 12-lead ECGs and basic demographics without relying on imaging, laboratory, or vital sign data. Our contributions are: (1) Predictive significance: prediction of a broad range of pulmonary findings from CXR data using ECGs with significant predictive capabilities. (2) Interpretability: explanations of model predictions to enhance trust, alongside comparison of feature importance with existing literature to validate findings and reveal novel links between cardiac and pulmonary systems. (3) Clinical relevance: Showing that ECGs, as a fast, non-invasive, and low-cost tool, can serve as a complementary approach to existing diagnostic pathways for pulmonary findings represents a step forward toward integrating ECG-derived insights into routine practice to enhance predictive capabilities and support earlier, more accessible detection. By the use of early, low-cost clinical data to approximate imaging outcomes, this work provides both practical and methodological advances at the intersection of machine learning, cardiology, and pulmonology.

\section{Background}\label{sec2}

\subsection{Pulmonary and cardiopulmonary conditions}

Pulmonary and cardiopulmonary conditions, such as acute respiratory distress syndrome (ARDS), pulmonary embolism, and heart failure, are among the leading causes of intensive care admissions and mortality \citep{ranieri2012ards,liu2022risk}. These diseases often present abruptly and progress rapidly, requiring immediate clinical intervention to prevent organ failure or death \citep{liu2018acute}. In critically ill patients, distinguishing overlapping symptoms such as dyspnea, hypoxia, and hemodynamic instability poses a significant diagnostic challenge. Early and accurate identification is essential for initiating appropriate therapy and improving outcomes \citep{arrive2021early}.

\subsection{Limitations of current diagnostic approaches}

Standard diagnostic tools including imaging (e.g., chest X-ray, CT), blood gas analysis, and invasive hemodynamic monitoring are valuable but often limited by logistics, delays, or patient instability \citep{jaccadv2025respfail}. Transporting critically ill patients for these diagnostic tests increases risk, while invasive procedures carry infection hazards \citep{gadre2018copd}. Moreover, such diagnostic tools do not allow for continuous monitoring at the bedside. Clinical signs may also be nonspecific or masked by sedation, mechanical ventilation, or comorbidities, making timely recognition of deterioration challenging. These challenges underscore the need for bedside, non-invasive, and continuous monitoring tools that support fast and reliable detection of critical conditions \citep{ceasovschih2024review}.

\subsection{ECG as potential diagnostic tool beyond cardiac conditions}

Recently, previous work has investigated the predictive performance of the ECG for diverse conditions beyond the cardiovascular system either with ECG features or ECG waveforms. From the ECG features side some of the conditions detected are neoplasms \citep{lopez2025explainable} as well as liver diseases conditions \citep{alcaraz2025electrocardiogram}, furthermore, the ECG features has been evaluated as discriminative factor in intensive care scenarios for mechanical ventilation as well as other clinical outcomes such as mortality and length of stay \citep{erdem2023correlation}. Similarly, from the ECG waveforms side, some examples are risk prediction of non-cardiac surgery using 12-leads \citep{harris2024risk}, pulmonay hypertension which reflect cardiac adaptation to a pulmonary vascular problem \citep{dubrock2024electrocardiogram}, as well as seizure detection \citep{mporas2015seizure}.

\section{Methods}\label{sec3}

\subsection{Dataset} 

\begin{figure*}[ht]
    \centering
    \includegraphics[width=1\linewidth]{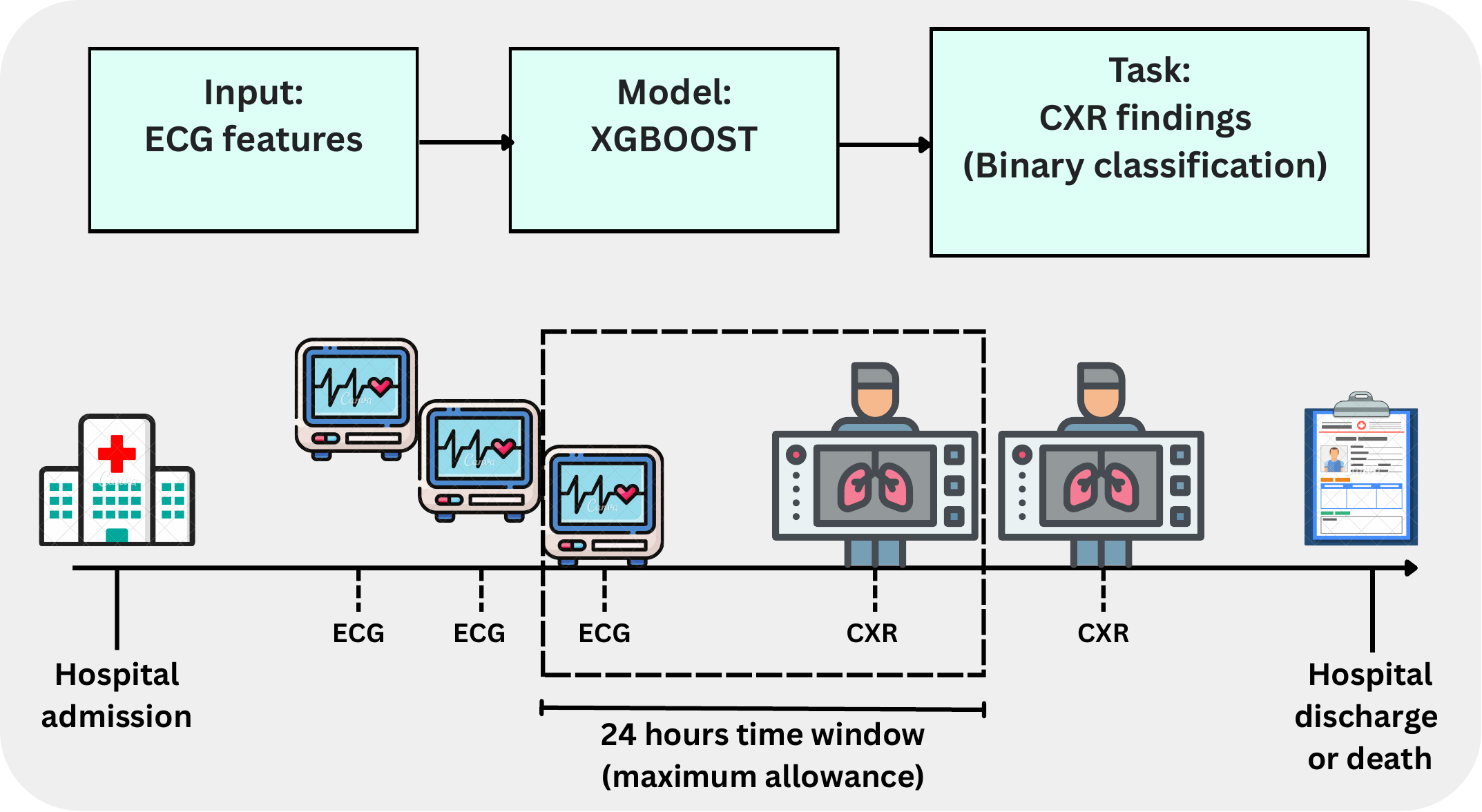}
    \caption{Overview of the sampling and modeling framework. Electrocardiograms (ECGs) from the MIMIC-IV-ECG dataset \citep{gow2023mimic} were temporally aligned with chest radiograph (CXR) findings from the CXR-LT dataset \citep{holste2024towards}, both aligned via the MIMIC-IV database \citep{johnson2023mimic}. Each CXR report was paired with the closest preceding ECG within a maximum tolerance of 24 hours. These paired samples were used to train an XGBoost model on structured ECG features for predicting radiographic findings. A stratified patient-level split (18:1:1) created training, validation, and test cohorts without patient overlap, ensuring balanced demographic and clinical distributions.}
    \label{fig:abstract}
\end{figure*}

Using complementary data sources, we paired ECGs from the MIMIC-IV-ECG dataset \citep{gow2023mimic} with radiographic labels from the CXR-LT dataset \citep{holste2024towards}, both integrated within the MIMIC-IV clinical database \citep{johnson2023mimic}. The MIMIC database was selected due to its large scale, rich demographic diversity, and clinical heterogeneity, supporting the development of robust and generalizable models. For model training and evaluation, we adopted a stratified patient-level split with an 18:1:1 ratio for training, validation, and test sets. This design avoided patient overlap across folds and preserved balanced distributions of age, sex, and diagnostic categories, following best practices outlined in \citep{strodthoff2024prospects}. To construct ECG–CXR pairs, we linked each chest radiograph interpretation, which is based on specialist reports in CXR-LT, with the temporally closest ECG recorded within the preceding 24 hours. This ensured that ECG features reflected the patient’s physiological state proximate to the imaging event, allowing us to explore the feasibility of predicting radiographic findings from routinely collected, non-invasive ECG data.

\subsection{Features}

\begin{table}[ht]
\caption{Descriptive statistics of the study population across training, validation, and test sets. Continuous variables are presented as median [IQR], categorical variables as count (\%).}
\label{tab:descriptive_total}
\centering
\small
\begin{tabular}{llll}
\toprule
Variable & Train (n = 83,953) & Val (n = 4,633) & Test (n = 4,633) \\
\midrule
\multicolumn{4}{c}{\textbf{Demographics}} \\
Age, years [IQR] & 66 [53-78] & 66 [54-78] & 66 [54-78] \\
\midrule
Age bins \\
18-53 & 21,177 (25.22\%) & 1,093 (23.59\%) & 1,093 (23.59\%) \\
53-66 & 21,797 (25.96\%) & 1,153 (24.89\%) & 1,153 (24.89\%) \\
66-78 & 20,845 (24.83\%) & 1,176 (25.38\%) & 1,176 (25.38\%) \\
78-99 & 20,134 (23.98\%) & 1,211 (26.14\%) & 1,211 (26.14\%) \\
\midrule
Female & 39,628 (47.20\%) & 2,188 (47.23\%) & 2,187 (47.20\%) \\
Male   & 44,325 (52.80\%) & 2,445 (52.77\%) & 2,446 (52.80\%) \\
\midrule
\multicolumn{4}{c}{\textbf{ECG features (median [IQR])}} \\
RR interval (ms) & 740 [612-870] & 740 [612-870] & 732 [612-869] \\
QRS axis (°) & 14 [-15-48] & 14 [-15-47] & 17 [-13-50.5] \\
P axis (°) & 52 [34-65] & 53 [33-66] & 53 [33-66] \\
T axis (°) & 43 [16-70] & 44 [16-70] & 44 [16-69] \\
P wave duration (ms) & 110 [96-120] & 110 [96-120] & 108 [96-118] \\
QRS duration (ms) & 93 [84-106] & 92 [84-104] & 92 [84-106] \\
PR segment (ms) & 156 [138-176] & 156 [138-176] & 156 [138-178] \\
QT segment (ms) & 384 [350-418] & 383 [350-418] & 384 [350-418.75] \\
QRS-T interval (ms) & 284 [254-316] & 286 [254-316] & 284.5 [254-314] \\
PT interval (ms) & 542 [500-586] & 542 [498-582] & 540 [500-584] \\
QTc (ms) & 445 [424-472] & 445 [423-470] & 445 [426-472] \\
P/RR ratio & 0.14 [0.12-0.17] & 0.14 [0.12-0.17] & 0.14 [0.12-0.17] \\
QRS/RR ratio & 0.13 [0.11-0.16] & 0.13 [0.11-0.16] & 0.13 [0.11-0.16] \\
QT/RR ratio & 0.53 [0.47-0.58] & 0.52 [0.47-0.58] & 0.53 [0.47-0.58] \\
PR/QT ratio & 0.41 [0.36-0.47] & 0.41 [0.36-0.46] & 0.41 [0.37-0.47] \\
P-QRS axis diff (°) & 34 [15-62] & 34 [15-60] & 33 [15-60] \\
QRS-T axis diff (°) & 43 [18-101] & 40 [17-96] & 44 [19-102] \\
P-T axis diff (°) & 25 [11-49] & 24 [11-45] & 25.5 [12-50] \\
\bottomrule
\end{tabular}
\end{table}

The original dataset contained nine ECG-derived measurements: QRS axis, T wave axis, P wave axis, P onset, P end, QRS onset, QRS end, T end, and RR interval. Implausible values outside physiologically valid ranges (e.g., QRS/T/P axis $<-360^\circ$ or $>360^\circ$, intervals $<0$ or $>5000$ ms) were set to missing. From these raw measurements, we engineered additional clinically meaningful features: PR interval, QRS duration, QT interval, and QTc (Bazett’s correction). Ratios such as P/RR, QRS/RR, QT/RR, and PR/QT were further derived to normalize wave durations by heart rate. Axis differences (P–QRS, QRS–T, and P–T) were also calculated to capture conduction and repolarization relationships. In total, this expanded feature set provided a richer representation of ECG morphology and timing. Demographic information was incorporated by adding patient age (both as a continuous variable and stratified into four bins: 18–53, 53–66, 66–78, and 78–99 years) and sex (male/female). Descriptive statistics for all features across training, validation, and test cohorts are provided in Table~\ref{tab:descriptive_total}.

During our experiments and after engineering additional features from the original nine ECG-derived variables, performance decreased for certain disease labels, likely due to redundancy, noise, or irrelevant information introduced by the expanded feature space. To address this, Recursive Feature Elimination (RFE) was employed as a systematic feature selection strategy. RFE iteratively removes the least important features based on model weights, allowing the algorithm to focus on the subset that maximizes predictive performance. Importantly, feature selection was conducted independently for each label, since the discriminative ECG characteristics vary across different diseases. This approach not only improved predictive accuracy by reducing overfitting and eliminating irrelevant variables, but also enhanced interpretability by highlighting label-specific ECG features with the greatest clinical relevance.

\subsection{Targets}
In this study, we aimed to predict 45 distinct chest X-ray findings, covering a wide spectrum of pulmonary, thoracic, and cardiopulmonary abnormalities. These targets span acute and chronic conditions, structural changes, degenerative features, and the presence of medical devices or interventions, providing a comprehensive representation of thoracic pathology. The investigated targets include bone and musculoskeletal issues (Osteopenia, Kyphosis, Scoliosis, Rib Fracture, Fracture), pulmonary parenchymal and airspace conditions (Emphysema, Bulla, Atelectasis, Consolidation, Lung Opacity, Lung Lesion, Nodule, Granuloma, Infiltration, Pneumonia, Fibrosis),  pleural and thoracic cavity conditions (Hydropneumothorax, Pleural Effusion, Pleural Other, Pleural Thickening, Pneumothorax, Pneumoperitoneum, Pneumomediastinum, Fissure), cardiovascular and mediastinal conditions (Calcification of the Aorta, Tortuous Aorta, Cardiomyopathy, Cardiomegaly, Enlarged Cardiomediastinum, Pulmonary Hypertension, Infarction),  abdominal or diaphragmatic conditions (Hernia),  medical devices and supports (Support Devices), Infectious and granulomatous conditions (Tuberculosis), as well as miscellaneous conditions (Adenopathy, Mass, Normal). Each target corresponds to a binary label indicating the presence or absence of the respective radiological finding, derived from CXR reports linked to the imaging studies.

\subsection{Models}
We employed gradient-boosted decision trees (XGBoost) as individual binary classifiers for each target. XGBoost was chosen for its robustness to outliers, its ability to handle missing data natively, avoiding preprocessing steps such as imputation that can distort feature distributions or introduce bias. This model is particularly well-suited for tabular data, which has been used a gold-standard method in the literature and still competitive against other models such as neural networks \citep{erickson2025tabarena}. For training hyperparameters, we set the maximum tree depth to 1, enabling the model to learn decision stumps that reduce overfitting. We used 1000 boosting rounds. The random seed was fixed at 42 for reproducibility, and the evaluation metric was the log loss.

\subsection{Performance Evaluation}
Model performance was assessed for each label using the Area Under the Receiver Operating Characteristic Curve (AUROC), with 95\% confidence intervals estimated via 1000 bootstrapping iterations on the test set to quantify uncertainty. AUROC was chosen because it evaluates overall ranking performance and is robust to class imbalance. Recent studies have also highlighted its advantages over alternative metrics, such as the area under the precision-recall curve (AUPRC), particularly in imbalanced settings \citep{mcdermott2024closer}. To evaluate model calibration, we generated calibration curves on the internal test set. For this, we applied a model-agnostic calibration approach, fitting isotonic regression models on the validation set and then reporting the calibrated results on the test set. Furthermore, we assessed clinical usefulness through decision curve analysis, benchmarking the net benefit of our model against standard reference strategies, namely “refer all” and “refer none” \cite{vickers2006decision}.

\subsection{Interpretability}
Although XGBoost provides inherent explanations such as feature importance, we employed SHapley Additive exPlanations (SHAP) \citep{lundberg2020local}, a game-theoretic approach that quantifies the contribution of each feature to individual predictions. SHAP was chosen because it provides consistent, locally accurate, and model-agnostic explanations, allowing us to interpret predictions at both the individual patient level and across the population. This approach enables identification of which ECG and demographic features exerted the strongest influence on the model’s output for each investigated condition, enhancing transparency and clinical trust in the predictive framework.

\section{Results}\label{sec4}

\subsection{Discriminative performance}

\begin{figure}[ht]
    \centering
    \includegraphics[height=0.80\textheight, width=0.80\linewidth, keepaspectratio]{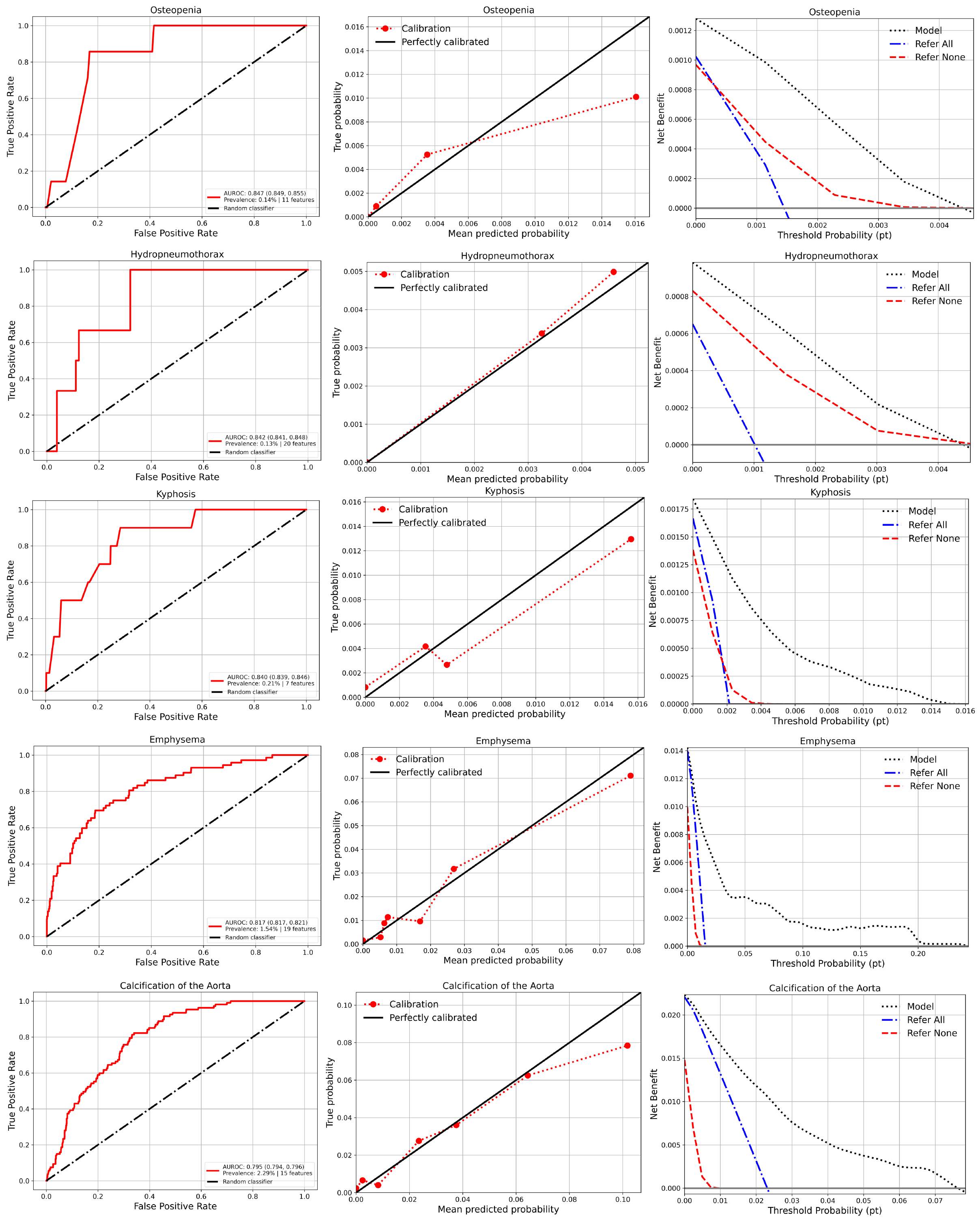}
    \caption{Discriminative performance for the top five predictive chest X-ray findings. Rows correspond to the labels Osteopenia, Hydropneumothorax, Kyphosis, Emphysema, and Calcification of the Aorta. Columns show AUROC with 95\% confidence intervals, calibration curves, and decision curve (net benefit) analysis. Calibration curves indicate good agreement between predicted probabilities and observed outcomes, with minimal average deviation. Decision curve analysis demonstrates that, across a wide range of threshold probabilities, the model (black dashed line) provides greater net benefit than reference strategies of treating all (blue dashed line) or treating none (red dashed line), highlighting potential clinical usefulness while avoiding unnecessary interventions.}
    \label{fig:results}
\end{figure}

Figure~\ref{fig:results} summarizes the predictive performance of the model for the five labels with the highest discrimination: Osteopenia (AUROC 0.847, 95\% CI 0.849–0.855), Hydropneumothorax (0.842, 0.841–0.848), Kyphosis (0.840, 0.839–0.846), Emphysema (0.817, 0.817–0.821), and Calcification of the Aorta (0.795, 0.794–0.796). These results indicate that ECG-derived features, combined with demographic information, can provide meaningful predictive signals for a diverse set of thoracic and cardiopulmonary conditions. Notably, the highest-performing labels include both musculoskeletal and pulmonary vascular conditions, suggesting that subtle cardiac electrical patterns may capture systemic physiological changes beyond classical cardiac diagnoses. Calibration curves for these labels demonstrate strong agreement between predicted probabilities and observed event frequencies. The small deviations observed indicate that the model’s probabilistic outputs are well-calibrated, which is crucial for clinical decision-making where predicted risk may guide interventions. Decision curve analysis further highlights the potential clinical utility of the model. Across a wide range of threshold probabilities, the model consistently provides a greater net benefit than baseline strategies of referring all patients or referring none. The “treat all” strategy declines sharply, underscoring the risks associated with unnecessary interventions, while the model achieves its greatest net benefit at lower decision thresholds. 

Additional results for the remaining chest X-ray findings are presented in the Supplementary Material. Beyond the top five predictive labels discussed in the main text (Osteopenia, Hydropneumothorax, Kyphosis, Emphysema, and Calcification of the Aorta), 11 more labels achieved AUROC values above 0.70, demonstrating strong discriminative performance. These include Support Devices (0.769), Cardiomyopathy (0.762), Hernia (0.759), Tortuous Aorta (0.754), Pleural Other (0.750), Bulla (0.740), Subcutaneous Emphysema (0.727), Edema (0.726), Pleural Effusion (0.724), Pulmonary Hypertension (0.721), and Cardiomegaly (0.704). The remaining 29 labels exhibited moderate to low predictive performance.

\subsection{Explainability}

\begin{figure}[H]
    \centering
    \includegraphics[height=0.70\textheight, width=0.70\linewidth, keepaspectratio]{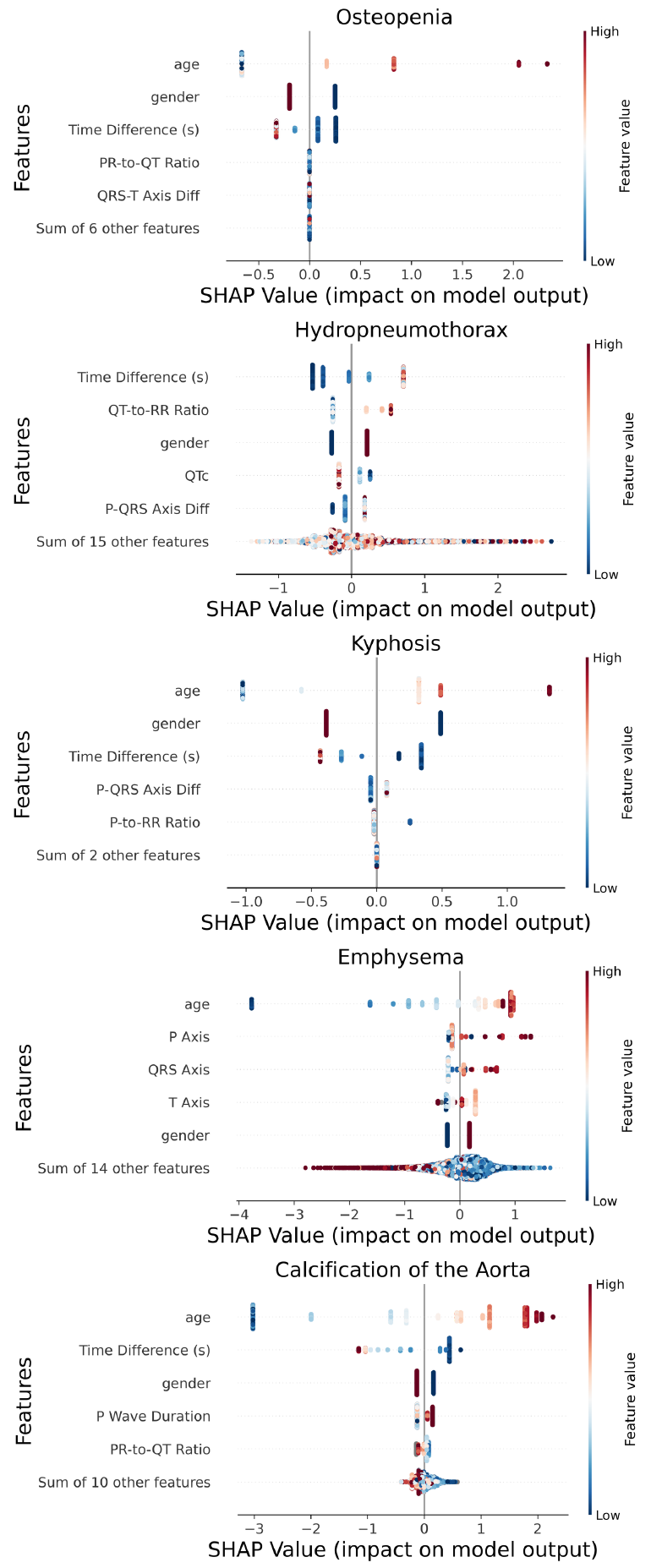}
    \caption{SHAP summary plots for a subset of target chest X-ray findings, illustrating the contribution of individual ECG features to model predictions. Top to bottom, left to right: Osteopenia—lower PR/QT ratio and smaller QRS-T axis difference drive predictions; Hydropneumothorax—higher QT/RR ratio, lower QTc, and increased P-QRS axis difference contribute positively; Kyphosis—larger P-QRS axis difference and lower P/RR ratio influence predictions; Emphysema—higher P, QRS, and T wave axes increase predicted risk; Calcification of the Aorta—longer P-wave duration and lower PR/QT ratio are associated with higher predicted probability. Colors indicate feature value, with red representing higher values and blue lower values.}
    \label{fig:shap}
\end{figure}

Figure \ref{fig:shap} contains the SHAP explanations on the five previously presented conditions. The SHAP analysis reveals that distinct ECG features drive predictions for different chest X-ray findings, highlighting the physiological specificity captured by the model. For instance, musculoskeletal and bone-related findings such as Osteopenia and Kyphosis are influenced by timing and axis ratios (PR/QT, P-to-RR, QRS-T axis differences), while pulmonary conditions like Emphysema and Hydropneumothorax are associated with amplitude and axis shifts (P, QRS, T axes, QT/RR ratio). Vascular-related conditions, such as Calcification of the Aorta, show contributions from conduction durations (P-wave) and interval ratios (PR/QT), suggesting that subtle alterations in cardiac electrical activity reflect systemic changes detectable in ECGs. Overall, these results demonstrate that the model not only predicts CXR findings accurately but also aligns with plausible physiological mechanisms, enhancing interpretability and clinical trust in the predictions.

\section{Discussion}\label{sec5}

\subsection{Summary of Findings}
While no individual label reached an AUROC above 0.9, the models demonstrated strong predictive performance for a substantial subset of labels. Nine labels achieved AUROC values above 0.75, and 16 exceeded 0.7, reflecting meaningful discriminative ability across a diverse set of chest X-ray findings. Calibration was generally robust, with confidence intervals indicating that predicted probabilities reliably reflected observed outcomes. Decision curve analysis further highlighted the potential clinical utility of the model, showing positive net benefit across most well-predicted labels and suggesting that these predictions could meaningfully inform real-world clinical decision-making. Importantly, by using SHAP values, our framework provides interpretable predictions, revealing which ECG features contribute most for each label. This transparency enhances clinical trust and supports further refinement and adoption of ECG-based predictive models in cardiopulmonary diagnostics.

\subsection{Clinical Utility}
The ECG-based predictive framework presented in this study has broad potential applications across multiple clinical scenarios. In time-sensitive settings such as the intensive care unit, rapid non-invasive predictions of chest X-ray findings could assist clinicians in triaging patients, identifying high-risk cases, and prioritizing imaging or interventions without delay. In routine diagnostic workflows, the model could serve as an adjunct to standard assessments, providing early warnings of pulmonary, musculoskeletal, or cardiopulmonary abnormalities and enabling more proactive patient management. Beyond diagnosis, predictions could inform treatment planning by highlighting systemic conditions or complications that might otherwise go undetected, supporting decisions such as ventilator adjustments, fluid management, or monitoring for pulmonary hypertension. Taken together, these capabilities suggest that ECG-derived predictive models could complement existing diagnostic and therapeutic strategies, ultimately improving timeliness, precision, and safety in patient care.

\subsection{Clinical Insights}
In this section, we focus on the three target labels with the highest discriminative performance (AUROC), examining the most influential features identified by SHAP analysis. For each label, we relate our observations to existing literature, highlighting both established and potentially novel associations.

\subsubsection{Osteopenia} 
Age emerged as the strongest predictor, alongside ECG-derived features including the PR-to-QT ratio and QRS-T axis difference. The link between age and bone density loss is well established, and low bone mineral density has been associated with increased cardiovascular risk, such as atherosclerosis and coronary artery disease \citep{pmc3258682}. These findings suggest that subtle cardiovascular changes, reflected in ECG parameters, may be indirectly related to bone mineral density reduction. To our knowledge, the specific association of PR-to-QT ratio or QRS-T axis difference with osteopenia has not been reported. We hypothesize that a lower PR-to-QT ratio may reflect subtle alterations in atrioventricular conduction related to systemic metabolic changes, while a smaller QRS-T axis difference could indicate modifications in ventricular repolarization linked to cardiovascular remodeling accompanying osteopenia.

\subsubsection{Hydropneumothorax}
The most influential ECG features were QT-to-RR ratio, QTc, and P-QRS axis difference. Pneumothorax and hydropneumothorax are known to induce ECG changes such as axis deviation, ST-segment shifts, T-wave alterations, and arrhythmias \citep{pneumothorax_review,pmc5385512}. Rightward axis deviation has been linked to mechanical displacement of the heart, and QRS amplitude changes often reverse after lung re-expansion \citep{pmc7989373}. However, the involvement of QT-to-RR ratio, QTc, and P-QRS axis difference has not been specifically documented. We hypothesize that an increased QT-to-RR ratio reflects compensatory changes in ventricular repolarization under altered intrathoracic pressure, while a larger P-QRS axis difference may result from spatial shifts in atrial versus ventricular depolarization due to lung collapse or effusion.

\subsubsection{Kyphosis}
Key predictive features included age, P-QRS axis difference, and P-to-RR ratio. Age is a well-known risk factor for degenerative spinal changes and kyphotic deformity. Previous studies have examined biomechanical and spinopelvic alterations associated with kyphosis \citep{kyphosis_frontiers}. Yet, no published work appears to investigate whether ECG features such as P-QRS axis difference or P-to-RR ratio are affected by spinal structural changes. We hypothesize that a larger P-QRS axis difference may reflect altered atrial-to-ventricular conduction due to changes in thoracic geometry, while a lower P-to-RR ratio could indicate subtle modifications in atrial depolarization timing relative to overall cardiac cycle length influenced by spinal curvature.

\subsection{Limitations and Future Work}
This study has several limitations. First, the lack of external validation is a key constraint: Models were developed and tested solely on the MIMIC-IV database, so their generalizability to other institutions and populations remains to be demonstrated \citep{ramspek2021external}. Second, we relied on pre-extracted ECG features rather than raw waveforms. While this facilitates interpretability and reduces computational demands, it may limit the model’s ability to capture complex signal patterns. Deep learning approaches on raw ECGs have shown potential for improved performance, albeit often with reduced interpretability \citep{alcaraz2025enhancing}. Third, incorporating additional clinical variables, such as laboratory measurements, could further enhance the predictive performance. Multimodal approaches combining ECG features with imaging data, such as a combination of chest X-rays itself, may also offer further improvements \citep{nagai2025enhanced}. 

\section{Statistical Results}\label{sec6}
The experiments were performed on single compute nodes equipped with 100GB of RAM and 24 CPU cores. These resources were provided by Carl von Ossietzky University of Oldenburg. All computational tasks were carried out using XGBoost version 3.0.5 and Python version 3.10.8.

\section*{Declarations}

\subsection*{Ethics approval and consent to participate}
This study used the publicly available MIMIC-IV-ECG and CXR-LT databases. The use of MIMIC-IV-ECG has received ethical approval from the Institutional Review Boards of the Massachusetts Institute of Technology (MIT) and Beth Israel Deaconess Medical Center (BIDMC), with a waiver of informed consent due to the use of de-identified data. Access to the database was granted to the authors after completion of the required training in human subjects research. The use of CXR-LT shared task uses image data from MIMIC-CXR-JPG v2.0.0 and generates labels from free-text radiology reports in MIMIC-CXR, a de-identified dataset that we gained access to through a PhysioNet Credentialed Health Data Use Agreement.

\subsection*{Consent for publication}
Not applicable.

\subsection*{Availability of data and materials}
The study used the publicly available MIMIC-IV-ECG \url{https://physionet.org/content/mimic-iv-ecg/1.0/} and the CXR-LT \url{https://physionet.org/content/cxr-lt-iccv-workshop-cvamd/2.0.0/} databases which can be accessed via PhysioNet after completion of the required data use agreement and training.  
Code for dataset preprocessing and experimental replication is available in our dedicated repository: \url{https://github.com/UOLMDA2025/CardioCXR}.

\subsection*{Competing interests} 
Upon manuscript submission, all authors completed the author disclosure form, confirming the absence of any conflicts of interest.

\subsection*{Funding}
This research received no specific grant from any funding agency in the public, commercial, or not-for-profit sectors.

\subsection*{Authors' contributions}
J.M.: Writing - original draft, Conceptualization, Methodology, Investigation, Data curation, Software, Validation, Visualization; O.Z.: Writing - original draft, Conceptualization, Methodology, Investigation, Data curation, Software, Validation, Visualization;  N.S.: Writing - review \& editing, Conceptualization, Methodology, Supervision, Project administration, Resources; J.M.L.A.: Writing - review \& editing, Conceptualization, Methodology, Supervision, Project administration, Resources.

\subsection*{Acknowledgements}
Not applicable.

\newpage

\begin{appendices}

\section{Additional Results}\label{secA1}

\begin{table}[!ht]
    \centering
    \small
    \begin{tabular}{lccc}
        \toprule
        \textbf{Label} & \textbf{AUROC} & \textbf{95\% CI lower} & \textbf{95\% CI upper} \\
        \midrule
        Osteopenia & 0.847 & 0.849 & 0.855 \\
        Hydropneumothorax & 0.842 & 0.841 & 0.848 \\
        Kyphosis & 0.840 & 0.839 & 0.846 \\
        Emphysema & 0.817 & 0.817 & 0.821 \\
        Calcification of the Aorta & 0.795 & 0.794 & 0.796 \\
        Support Devices & 0.769& 0.769& 0.769\\
        Cardiomyopathy & 0.762& 0.762& 0.772\\
        Hernia & 0.759 & 0.755 & 0.758 \\
        Tortuous Aorta & 0.754 & 0.752 & 0.755 \\
        Pleural Other & 0.750 & 0.745 & 0.751 \\
        Bulla & 0.740& 0.733 & 0.746 \\
        Subcutaneous Emphysema & 0.727 & 0.724 & 0.731 \\
        Edema & 0.726 & 0.724 & 0.725 \\
        Pleural Effusion & 0.724 & 0.724 & 0.725 \\
        Pulmonary Hypertension & 0.721 & 0.721 & 0.730 \\
        Cardiomegaly & 0.704 & 0.703 & 0.704 \\
        Normal & 0.693 & 0.692 & 0.694 \\
        Pneumoperitoneum & 0.685 & 0.683 & 0.690 \\
        Lung Lesion & 0.684 & 0.683 & 0.689 \\
        Nodule & 0.681 & 0.679 & 0.681 \\
        Pneumothorax & 0.675 & 0.675 & 0.677 \\
        Atelectasis & 0.659 & 0.658 & 0.659 \\
        Hilum & 0.648 & 0.648 & 0.649 \\
        Granuloma & 0.637 & 0.636 & 0.640 \\
        Consolidation & 0.629 & 0.628 & 0.630 \\
        Lung Opacity & 0.629 & 0.629 & 0.630 \\
        Adenopathy & 0.628 & 0.626 & 0.630 \\
        Pulmonary Embolism & 0.628 & 0.626 & 0.632 \\
        Fissure & 0.624 & 0.622 & 0.628 \\
        Rib Fracture & 0.584 & 0.581 & 0.584 \\
        Mass & 0.577 & 0.575 & 0.579 \\
        Pleural Thickening & 0.574 & 0.571 & 0.577 \\
        Infarction & 0.574 & 0.567 & 0.575 \\
        Pneumomediastinum & 0.569 & 0.562 & 0.570 \\
        Pneumonia & 0.565 & 0.564 & 0.566 \\
        Infiltration & 0.566 & 0.564 & 0.567 \\
        Fracture & 0.554 & 0.553 & 0.555 \\
        Enlarged Cardiomediastinum
& 0.539& 0.539& 0.540\\
        Scoliosis& 0.533& 0.532& 0.535\\
        Tuberculosis & 0.499 & 0.501 & 0.511 \\
        Fibrosis & 0.468 & 0.468 & 0.476 \\
        \bottomrule
    \end{tabular}
    \caption{AUROC values with 95\% confidence intervals for all predicted labels.}
    \label{tab:predictive_performance}
\end{table}

\end{appendices}

\newpage

\bibliography{sn-bibliography}

\end{document}